\documentclass[aps,showpacs,twocolumn,preprintnumbers,amsmath,amssymb,superscriptaddress]{revtex4}
\usepackage{graphicx}
\begin{document}
\title{High Pressure studies of the magnetic phase transition in MnSi: revisited}

\author{Alla E. Petrova}
\affiliation{Institute for High Pressure Physics of Russian
Academy of Sciences, Troitsk, Moscow Region, Russia}
\author{Vladimir N. Krasnorussky}
\affiliation{Institute for High Pressure Physics of Russian
Academy of Sciences, Troitsk, Moscow Region, Russia}
\author{T. A. Lograsso}
\affiliation{Ames Laboratory, Iowa State University, Ames, IA
50011, USA}
\author{Sergei M. Stishov}
\email{sergei@hppi.troitsk.ru} \affiliation{Institute for High
Pressure Physics of Russian Academy of Sciences, Troitsk, Moscow
Region, Russia}

\date{\today}

\begin{abstract}
New measurements of AC magnetic susceptibility and DC resistivity of a
high quality single crystal MnSi were carried out at high pressure
making use of helium as a pressure medium. The form of the AC magnetic
susceptibility curves at the magnetic phase transition suddenly changes
upon helium solidification. This implies strong sensitivity of magnetic
properties of MnSi to non hydrostatic stresses and suggests that the
early claims on the existence of a tricritical point at the phase
transition line are probably a result of misinterpretation of the
experimental data. At the same time resistivity behavior at the phase
transition does not show such a significant influence of helium
solidification. The sharp peak at the temperature derivative of
resistivity, signifying the first order nature of the phase transition
in MnSi successfully survived helium crystallization and continued the
same way to the highest pressure.
\end{abstract}


\maketitle
As it was found long ago, the intermetallic compound MnSi acquired a
long period helical magnetic structure at T${\approx}$ 29 K \cite{1,2}.
Itinerant nature of magnetism in MnSi was established in \cite{3}. First
experiments on the influence of high pressure on the phase transition
in MnSi showed that the transition temperature decreased with pressure
and tended to zero at about 1.4 GPa \cite{4}. This feature of MnSi promised
the opportunity of observation of quantum critical behavior.

Since then, quite a number of papers has been devoting to high pressure
studies of the phase transition in MnSi (see for instance \cite{5,6,7,8,9,10,11,12,13,14,15}).
Among them one should point at a paper \cite{5}(see also \cite{9}) where authors
claimed the discovery of a tricritical point at the phase transition
line of MnSi, based on the evolution of the AC magnetic susceptibility
of MnSi ($\chi_{AC}$) with pressure. Later on a theory
was developed that declared generic \ nature of first order character
of phase transitions in ferromagnets at low temperatures \cite{6}.

Non-Fermi liquid behavior has been observed in an extended region of
pressure above the quantum critical point \cite{11}. The specific magnetic
structure (partial order) of the non-Fermi-liquid phase of MnSi was
described in Ref. \cite{12}. Phase inhomogeneity
in the region surrounding the transition line from about 10 K to the
lowest temperatures was reported in Ref. \cite{13,14}. Incidentally,
studies of the AC magnetic susceptibility of MnSi at high pressure \cite{15}
using fluid and solid helium as a pressure medium showed somewhat
different results from the data \cite{5}. In particular authors \cite{15}
concluded that the radical change of the AC magnetic susceptibility of
MnSi with pressure, observed in Ref. \cite{5}, could be influenced by
non-hydrostatic stresses, developing in a frozen pressure medium.
Precise lattice constant measurements at high pressure seemingly
indicate that the phase transition in MnSi is first order \cite{16}.

Meanwhile new studies of thermodynamic and transport properties of a
high quality single crystal of MnSi at ambient pressure suggested also
a first order nature of the corresponding magnetic phase transition
\cite{17,18,19} that again questioned the early proposed phase diagram \cite{5,6,7,8,9}.

Having a high quality single crystal of MnSi, which reveals sharp
features of the phase transition, never so clearly observed before, it
was appealing to conduct new studies of the phase diagram of MnSi with
helium as a pressure medium.

In this paper we report results of new measurements of resistivity and
AC magnetic susceptibility of MnSi at high pressure, created by
compressed helium.

The samples for the current measurements as well as
for a earlier study \cite{17} were cut from a single crystal of MnSi, grown
from melt by the Bridgman method. The resistivity was measured by a
four-terminal DC method. Measurements of the AC magnetic
susceptibility were carried out with a standard modulation technique at
a modulation frequency of 19 Hz. Temperature was measured by a
calibrated Cernox sensor imbedded in the cell body. Accuracy of the
Cernox sensor in the temperature range under study is about 0.02 K.
Pressure was measured by a calibrated manganin gauge while helium was
still in the fluid phase. Pressure in solid helium was calculated using
its known EOS. Details of the experimental techniques are described in
Ref. \cite{15,20,21,22}.

\begin{figure}[htb]
\includegraphics[width=80mm]{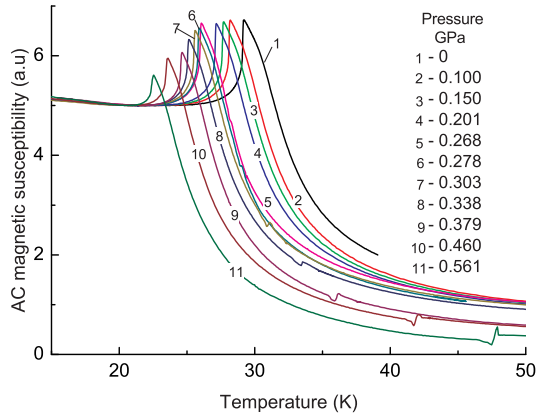}
\caption{\label{fig1} (Color online) Influence of pressure on behavior of the AC magnetic
susceptibility ($\chi_{AC}$) at the phase transition in
MnSi. Zigzag features in some of the curves are results of the helium
crystallization. }
\end{figure}

Results of the experimental studies are illustrated in Fig. \ref{fig1}{--}\ref{fig6}.

\begin{figure}[htb]
\includegraphics[width=80mm]{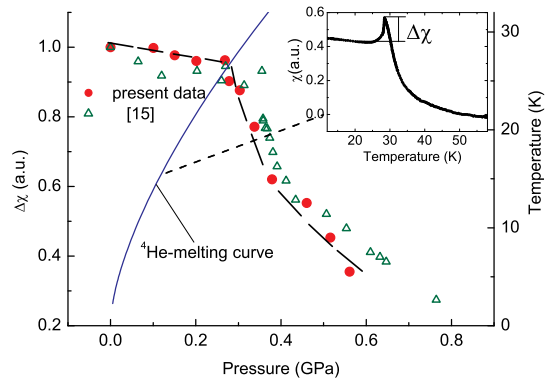}
\caption{\label{fig2} (Color online) Evolution of the form of $\chi_{AC}$ at
the phase transition in MnSi. The dash line corresponds to the
hcp{}-fcc transition in solid helium. The definition of
$\Delta\chi$ is given in the inset. }
\end{figure}

\begin{figure}[htb]
\includegraphics[width=80mm]{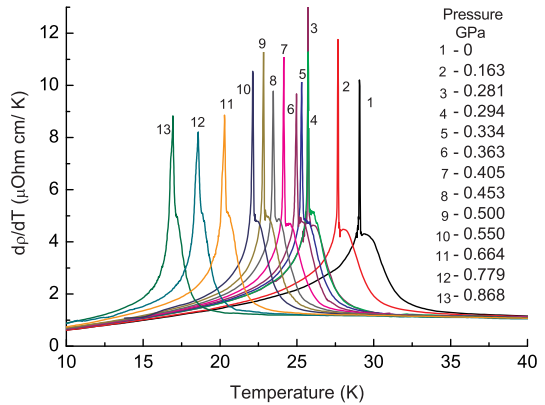}
\caption{\label{fig3} (Color online) Temperature dependence of derivative of resistivity of
MnSi (d${\rho}$/dT) at different pressures. The melting curve of helium
and the phase transition line in MnSi cross at 0.27 GPa. }
\end{figure}
As is seen in Figs. \ref{fig1} and \ref{fig2}, the sharp maximum of AC magnetic
susceptibility $\chi_{AC}$ does not change much up to
pressure of about 0.3 GPa, then starts to decrease rapidly at higher
pressures, as observed before \cite{15}. But one could notice subtle
differences between present data and the data \cite{15} (see Fig. \ref{fig2}). As it
is evident from Fig. \ref{fig2}, a drastic change of the form of the magnetic
susceptibility curve $\chi_{AC}$ in the current data
closely coincides with the helium melting point that clearly indicates
the effect of non hydrostatic stresses arising on helium
solidification. Slightly different behavior of
$\chi_{AC}$ observed in \cite{15}, most probably is
connected with a smaller size of the sample used in the old
measurements compare with the new one. So the non-hydrostatic stress
amplitude could reach a critical value only at some distance from the
melting point inside the solid helium domain. This situation led to
ignoring possible non-hydrostatic effects in the experiments \cite{15}
with subsequent misinterpretation of the experimental data. The new
measurements unambiguously show that the striking change of the
magnetic susceptibility curve is a result of non-hydrostatic stresses
in pressure media and has nothing to do with a change of character of
the phase transition in MnSi \cite{5,15,23}.

\begin{figure}[htb]
\includegraphics[width=80mm]{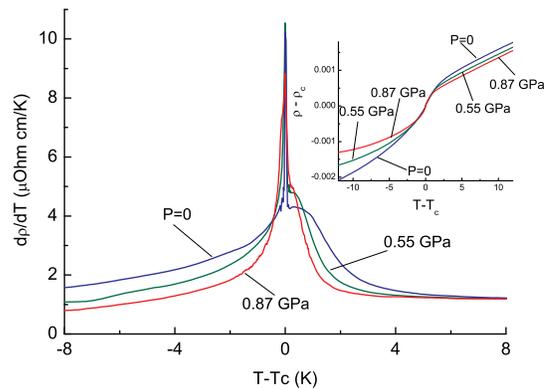}
\caption{\label{fig4} (Color online) Narrowing of the global maximum of the temperature
dependence of resistivity of MnSi (d$\rho/dT$) at high pressures. The
inset illustrates the nature of narrowing of d$\rho/dT$.
$T_{c}$ and $\rho_{c}$ are the phase
transition temperature and the value of resistivity at the transition
point. }
\end{figure}

\begin{figure}[htb]
\includegraphics[width=80mm]{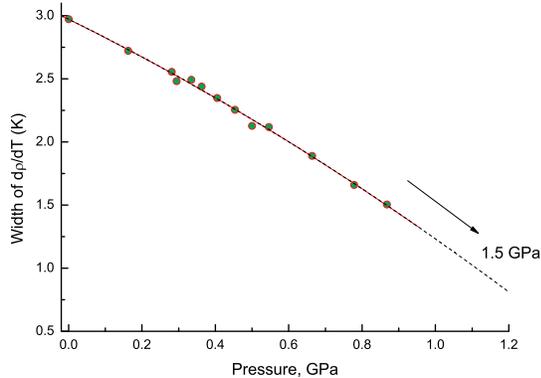}
\caption{\label{fig5} (Color online) Pressure dependence of the width of the global maximum
in d$\rho/dT$. The width is taken at the level of 3 ${\mu}$Ohm cm/K (see Fig. 3). Note
that the melting curve of helium and the phase transition line are
crossed at about 0.27 GPa. }
\end{figure}

\begin{figure}[htb]
\includegraphics[width=80mm]{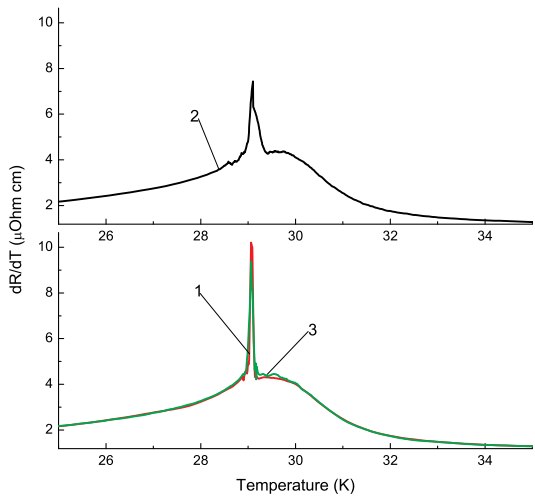}
\caption{\label{fig6} (Color online) Influence of fast pressure releasing on the peak of
d${\rho}/dT$ at the phase transition in MnSi. 1 {--} ambient pressure, 2
{--} ambient pressure after loading the sample to 1.2 GPa, 3 {--}
\ after holding the sample, experienced 1.2 GPa, for a week at ambient
pressure. }
\end{figure}

 Now we turn to Fig. \ref{fig3} where temperature derivatives of resistivity
d${\rho}$/dT in the vicinity of the phase transition are depicted as a
function of pressure. As one can see, d${\rho}$/dT does not experience
so dramatic a change across the helium melting line as the AC magnetic
susceptibility does. However, two obvious trends are seen. Both of them
are better illustrated in Fig.\ref{fig4}. The first trend is broadening the
sharp peaks of d${\rho}$/dT that become obvious at a pressure of 0.66
GPa. As was discussed earlier \cite{17}, sharp peaks of d${\rho}$/dT
originate from the first order nature of the phase transition in MnSi.
Hence, broadening of the peaks in d${\rho}$/dT at high pressure implies
smearing the phase transition by non-hydrostatic stresses. The second
one is narrowing of the global anomaly in d${\rho}$/dT, which
accompanies the phase transition and reveals itself as a satellite
rounded peak on the high temperature side of the phase transition. Note
that the d${\rho}$/dT anomaly scales perfectly with the corresponding
anomalies in heat capacity and thermal expansion coefficient data of
MnSi \cite{17,19}. Fig. \ref{fig5} shows an evolution of the width of maxima in
d${\rho}$/dT on pressure increasing. Surprisingly, in contrast with the
case of magnetic susceptibility, no trace of the helium crystallization
is seen on the corresponding curve. As it follows from Fig. \ref{fig3} and \ref{fig4}
narrowing of the anomaly signifies its general reduction despite the
fact that its amplitude slightly increases \cite{24}. More specifically this
implies decreasing the general abundance of spin fluctuations along the
transition line since electron scattering on spin fluctuations provides
a major contribution to the resistivity of MnSi around the phase
transition. Simple extrapolation of the curve in Fig. \ref{fig5} to the zero
width leads to the pressure value about 1.5 GPa, which almost exactly
corresponds to the phase transition pressure at T=0, as should be
expected.

Another effect observed in the current study of resistivity of MnSi is
defect generation upon pressure releasing, as is illustrated in Fig. \ref{fig6}.
It is seen that, after decreasing pressure from $\sim1.2\ GPa$ to zero in a
time span of less then one hour, the peak of d${\rho}$/dT appears to be
highly distorted. Remarkably, after holding the sample for a week at
ambient pressure and room temperature, the form of the peak completely
recovered its initial shape. All that did not happen when the sample
was loaded by 0.7 GPa of helium pressure. This observation tells us
that at some pressure, obviously more then 0.7 GPa, MnSi experiences
some sort of irreversible change on a short time scale that facilitates
generation of defects on pressure releasing.

Despite all these complications and even ignoring the data obtained
above 0.7 GPa, we are still able to derive certain conclusions in
regard to the phase transition in MnSi at high pressure.

First of all, nonhydrostatic stresses arising in solid helium strongly
influence the form of AC magnetic susceptibility of MnSi and the latter
can not serve as an indicator of the type of phase transition. So,
claims of the existence of a tricritical point on the phase
transition line in MnSi seem to be a result of misinterpretation of the
experimental data \cite{5,10,15}. Second, the electrical resistivity of
MnSi appears to be almost insensitive to the non{}-hydrostatic stresses
and the persistent sharp peak of d${\rho}$/dT experiences only slight
broadening with pressure. Third, the existence of d${\rho}$/dT peaks
at the highest pressures provides evidence that the magnetic phase
transition in MnSi continues to be first order in the whole pressure
range studied and most probably will stay the same way with further
increase of pressure.

 Finally, the different sensitivity of AC magnetic susceptibility and
electrical resistivity to non{}-hydrostatic environment probably
indicates the dual role of the spin fluctuations involved in the
corresponding physics. But detailed analysis of the situation should
wait until there is a proper understanding of the nature of the
satellite rounded maxima in d${\rho}$/dT as well as the heat capacity
and the thermal expansion coefficient at the phase transition in MnSi.

Authors are thankful  to  J.D. Thompson for reading the manuscript and valuable remarks.
TAL wish to acknowledge the support of the U.S. Department of Energy,
Basic Energy Sciences. AEP, SMS and VNK   appreciate support of the
Russian Foundation for Basic Research , Program
of the Physics Department of RAS on Strongly Correlated Systems
and Program of the Presidium of RAS on Physics of Strongly
Compressed Matter.


\begin{thebibliography}{99}
\bibitem{23} Holding the cell at 35 K and at 0.5 GPa pressure of solid helium for
more than 12 hours in attempt to reduce non hydrostatic stresses did
not effect the magnetic susceptibility of the sample.
\end{thebibliography}
\end{document}